\documentclass[10pt,journal,twocolumn]{IEEEtran}
\IEEEoverridecommandlockouts
\usepackage{amsmath,amsfonts,amssymb,mathtools}
\usepackage{cite}
\usepackage{xcolor}
\usepackage{graphicx}
\usepackage{array}
\usepackage{enumitem}
\setlist[itemize]{topsep=2pt, parsep=0pt, partopsep=0pt, leftmargin=*}
\usepackage{microtype,url}
\usepackage[font=footnotesize]{caption}
\usepackage{pgfplots}
\pgfplotsset{compat=1.17}

\newcommand{\Kstar}{K^{\star}}
\newcommand{\thmin}{\theta_{\min}}
\newcommand{\RE}{R_{\mathrm{E}}}

\title{Downlink MU-MIMO for LEO Direct-to-Device:\\ Feasibility, Architecture, and Implementation}

\author{Bikshapathi~Gouda, Neng~Wang, Masoud Sajadieh, and Meghna Agrawal
\thanks{B. Gouda, and N. Wang are with Skylo Technologies Europe Oy, Espoo, Finland (e-mail: \{bikshapathi, neng\}@skylo.tech).}
\thanks{M. Sajadieh, and M. Agrawal are with Skylo Technologies Inc., CA, USA (e-mail: \{masoud, meghna\}@skylo.tech).}}

\begin{document}
\maketitle
\begin{abstract}
Direct-to-device connectivity from low Earth orbit (LEO) satellites is commercially available, but its throughput is limited by the satellite antenna size and by payloads that separate users only by beams and frequency reuse. This article shows that a single LEO satellite with a meter-scale array can serve hundreds of users on the same time-frequency resource through multi-user multiple-input multiple-output precoding. The number of users it can separate with a single stream each, i.e., its spatial degrees of freedom (DoF), follows from the aperture in wavelengths, the orbit altitude, and the minimum elevation angle. With minimum mean-square error (MMSE) precoding, the throughput of these users grows with the signal-to-noise ratio (SNR), whereas maximum-ratio transmission becomes interference-limited. The precoder can be computed from the user positions and the satellite ephemeris and attitude, with a throughput loss of less than 0.5\% at 20~dB SNR compared with perfect channel knowledge under the 3GPP non-terrestrial channel models. The DoF also set the number of digital chains of a hybrid payload, and the functional split among satellites and ground trades inter-satellite-link and feeder-link load against on-board processing. As the satellite moves, the precoder can be updated from the known geometry every radio frame, i.e., 10~ms, while the computationally intensive user pairing can be kept for a few seconds with a throughput loss below 1\%. Finally, we keep the cell-specific signals on fixed beams while precoding the user data jointly across all cells of the satellite. For a $50\times50$ array at 600~km and a minimum elevation angle of $60^\circ$, one satellite serves 410 users at 6.1~bps/Hz on average, more than four times the throughput of fixed beams with four-color frequency reuse.

\end{abstract}

\section{Introduction}
\label{sec:intro}
A few years ago, connecting an ordinary smartphone directly to a satellite was considered impractical, whereas today it is a commercial reality. With the introduction of non-terrestrial networks (NTN) in Release~17 of the 3rd Generation Partnership Project (3GPP), NTN-capable smartphones and Internet of Things (IoT) devices attach to satellites using the same protocols they use on the ground~\cite{Lin21,TR38811}. Satellites in low Earth orbit (LEO) have become the preferred platform for this direct-to-device (D2D) service because of their short distance to the Earth, promising coverage everywhere with the phone already in the user's pocket. 

The throughput of LEO D2D services is limited by the user device, which typically has an omnidirectional antenna and a transmit power of a fraction of a watt, and is hundreds of kilometers away from the satellite. To close this link, D2D satellites carry very large antenna arrays, now reaching several tens to hundreds of square meters~\cite{AST}. However, these arrays are operated as conventional multibeam arrays, in which the coverage area is illuminated by a fixed grid of narrow beams, and neighboring beams use different frequencies to avoid mutual interference. Each beam therefore uses only a fraction of the spectrum and serves one user per time-frequency resource, so that the number of simultaneously served users is limited by the beam grid and the reuse pattern rather than by the number of users the array could separate spatially.

More than a decade ago, terrestrial networks faced the same bottleneck, where the number of simultaneously served users was limited by the number of cells and their reuse pattern, and higher throughput required more cells or more spectrum. Massive multiple-input multiple-output (MIMO) removed this bottleneck, since a base station with a large array serves many users with nearly orthogonal channels on the same time-frequency resource~\cite{Rusek13,Larsson14}. For LEO satellites, massive MIMO has been studied with precoding based on statistical channel knowledge~\cite{You20}, with hybrid beamforming to reduce the number of digital chains~\cite{Le26HBF}, and with precoding from reported user positions~\cite{Gui22}. Further, interference-aware precoding has been shown to allow aggressive frequency reuse and to increase the throughput of geostationary satellites~\cite{Vaz16}, and can also be applied to LEO satellites. These works address individual components, but not the limit on the number of users served by a single satellite. A common view is that a single satellite offers little spatial multiplexing, since users seen under closely spaced angles are not distinguishable with a small array. Therefore, additional throughput is sought from several cooperating satellites~\cite{Multisat24}. However, a sufficiently large array can separate such users.

This article revisits this common view and asks \emph{whether a single LEO satellite can serve hundreds of users on the same time-frequency resource, i.e., with frequency reuse~1, and what it takes}. We show that it can, since the D2D link is almost purely line-of-sight (LoS), the spatial signature of each user's channel is set by its direction from the satellite, and a large array can separate users whose directions differ by more than its beam width. Therefore, the design of a single-satellite multi-user MIMO (MU-MIMO) system follows from the geometry of the satellite-to-user link. From this geometry, we derive the spatial degrees of freedom (DoF) of a coverage area, i.e., the number of single stream users that one satellite can separate on the same resource, and show that with minimum mean-square error (MMSE) precoding, the throughput of these users grows with the signal-to-noise ratio (SNR), whereas maximum-ratio transmission (MRT) becomes interference-limited. Moreover, the user position suffices for precoding, even without a direct path. The DoF also set the number of digital chains of a hybrid payload, and based on the functional split and an on-board precoder design, we quantify the traffic load of the feeder link (FL), where the same traffic can also be relayed over inter-satellite links (ISLs) when no gateway is in view. Although the satellite geometry changes continuously, the user pairing can be fixed for seconds, while the precoder is updated every radio frame from the known geometry. Finally, for the multiple 5G New Radio (NR) cells of one satellite, we propose a cellular control plane with a cell-free data plane, which is transparent to the user and increases the throughput over per-cell operation. In our case study, a $50\times50$ array serves 410 users on the same 5~MHz carrier at 6.1~bps/Hz per user, i.e., 12.5~Gbps in total.

\begin{figure}[t]
\centering
\includegraphics[width=1\columnwidth]{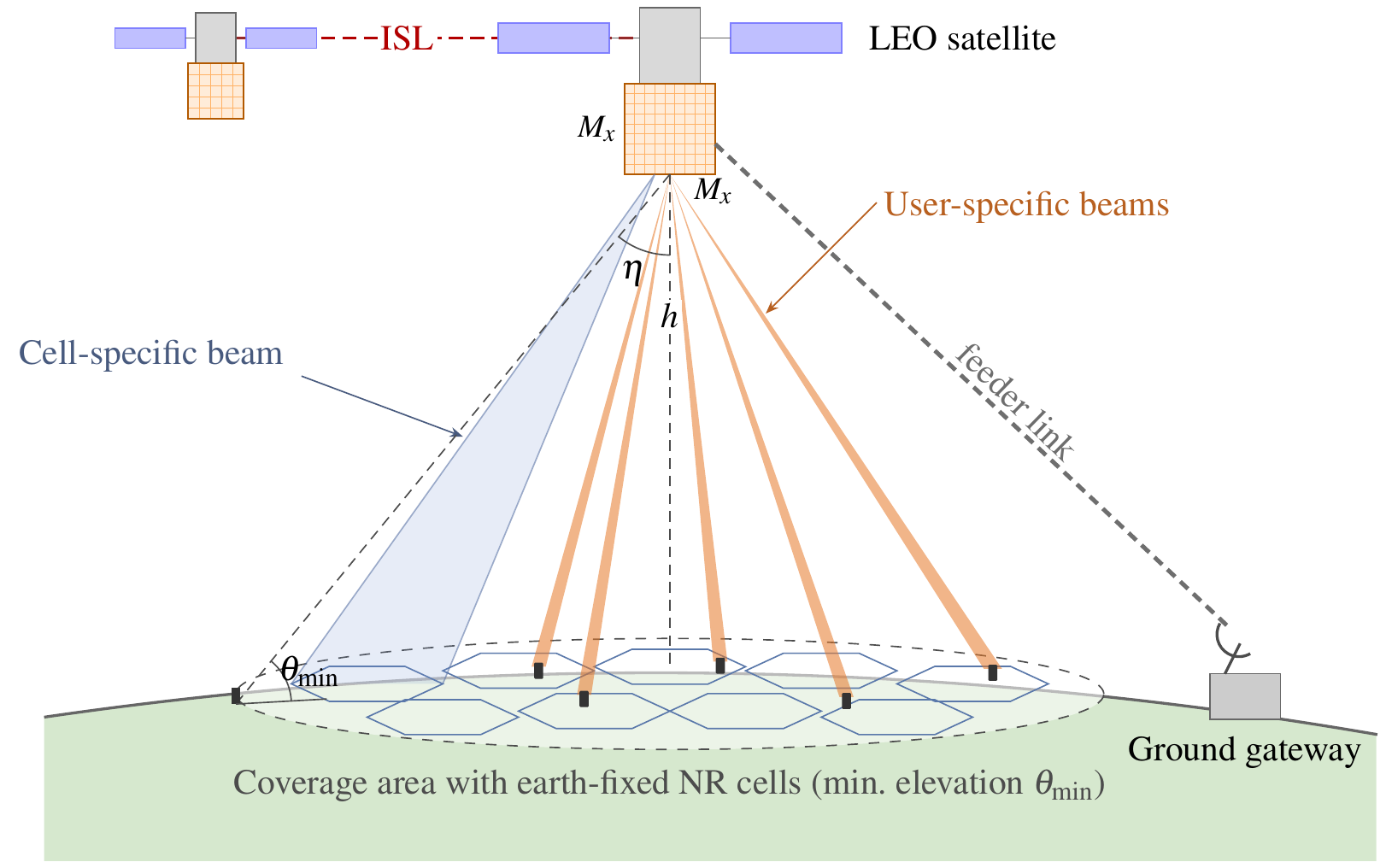}
\caption{Single-satellite MU-MIMO for LEO D2D, with cell-specific beams for the earth-fixed NR cells and user-specific beams for the user data.}
\label{fig:system}
\end{figure}

Fig.~\ref{fig:system} illustrates the considered LEO D2D system. The coverage area of the satellite is divided into earth-fixed NR cells, each illuminated by a cell-specific beam that carries the synchronization and system information needed for cell search and access, whereas the data of each user is delivered on its own user-specific beam, so that many users in the same or in neighboring cells share the same spectrum. The satellite is connected to the core network through an FL to a ground gateway, and to neighboring satellites through ISLs. In the case study, we use a $50\times50$ uniform planar array (UPA) with 2500 elements, about 3.4~m across, operating at 2.2~GHz from an altitude of 600~km. Unless stated otherwise, we assume full-buffer traffic, equal power per user with an SNR of 20~dB at nadir, ideal array calibration, and linear power amplifiers, so that the results indicate the achievable performance.

\section{Feasibility: How Many Users Can One Satellite Serve?}
\label{sec:feas}

\begin{figure*}[!t]
\centering
\pgfplotsset{fig2style/.style={
	width=6.4cm, height=5.5cm,
	xlabel near ticks, ylabel near ticks,
	x label style={font=\footnotesize}, y label style={font=\footnotesize},
	ticklabel style={font=\footnotesize},
	legend cell align=left,
	legend style={font=\tiny, inner sep=1pt, row sep=-1pt, fill opacity=0.75, draw opacity=1, text opacity=1},
	grid=both, grid style={gray!20},
}}
\begin{tikzpicture}
\begin{axis}[fig2style,
	xmin=0.5, xmax=6, ymin=1, ymax=1e4, ymode=log,
	xlabel={Aperture side length [m]},
	ylabel={Spatial DoF $K^\star$},
	legend style={at={(0.97,0.03)}, anchor=south east},
]
\addplot[dashdotted, line width=1pt, black]
table[x=D, y=horizon, col sep=comma] {Figs/fig2a_dof_lines.txt};
\addlegendentry{Horizon ($\theta_{\min}=0^\circ$)};
\addplot[line width=1pt, red]
table[x=D, y=th20, col sep=comma] {Figs/fig2a_dof_lines.txt};
\addlegendentry{$\theta_{\min}=20^\circ$};
\addplot[line width=1pt, teal]
table[x=D, y=th40, col sep=comma] {Figs/fig2a_dof_lines.txt};
\addlegendentry{$\theta_{\min}=40^\circ$};
\addplot[line width=1pt, blue]
table[x=D, y=th60, col sep=comma] {Figs/fig2a_dof_lines.txt};
\addlegendentry{$\theta_{\min}=60^\circ$};
\addplot[only marks, red, mark=x, mark size=2.5pt, forget plot]
table[x=D, y=th20, col sep=comma] {Figs/fig2a_dof_count.txt};
\addplot[only marks, teal, mark=x, mark size=2.5pt, forget plot]
table[x=D, y=th40, col sep=comma] {Figs/fig2a_dof_count.txt};
\addplot[only marks, blue, mark=x, mark size=2.5pt]
table[x=D, y=th60, col sep=comma] {Figs/fig2a_dof_count.txt};
\addlegendentry{Exact beam count};
\draw[densely dotted, gray] (axis cs:3.41,1) -- (axis cs:3.41,1e4);
\end{axis}
\node[font=\footnotesize, anchor=north] at (current axis.outer south) {(a)};
\end{tikzpicture}\hfill
\begin{tikzpicture}
\begin{axis}[fig2style,
	xmin=-10, xmax=20, ymin=0, ymax=7.5,
	xlabel={SNR at nadir [dB]},
	ylabel={Avg. rate per user [bps/Hz]},
	legend style={at={(0.03,0.97)}, anchor=north west},
]
\addplot[dashed, line width=1pt, black]
table[x=snr, y=ub, col sep=comma] {Figs/fig2b_los.txt};
\addlegendentry{Interference-free bound};
\addplot[line width=1pt, blue]
table[x=snr, y=mmse_cf, col sep=comma] {Figs/fig2b_los.txt};
\addlegendentry{MMSE, closed form};
\addplot[only marks, blue, mark=o]
table[x=snr, y=mmse, col sep=comma] {Figs/fig2b_los.txt};
\addlegendentry{MMSE, simulation};
\addplot[line width=1pt, red]
table[x=snr, y=mrt_cf, col sep=comma] {Figs/fig2b_los.txt};
\addlegendentry{MRT, closed form};
\addplot[only marks, red, mark=square]
table[x=snr, y=mrt, col sep=comma] {Figs/fig2b_los.txt};
\addlegendentry{MRT, simulation};
\draw[densely dotted, gray] (axis cs:8,0) -- (axis cs:8,7.5);
\node[font=\tiny, anchor=south west] at (axis cs:8.3,0.1) {$\rho=1/\epsilon$};
\end{axis}
\node[font=\footnotesize, anchor=north] at (current axis.outer south) {(b)};
\end{tikzpicture}\hfill
\begin{tikzpicture}
\begin{axis}[fig2style,
	xmin=-10, xmax=20, ymin=0, ymax=7.5,
	xlabel={SNR at nadir [dB]},
	ylabel={Avg. rate per user [bps/Hz]},
	legend style={at={(0.03,0.97)}, anchor=north west},
]
\addplot[line width=1pt, blue]
table[x=snr, y=mmse_tdlc, col sep=comma] {Figs/fig2c_tdl.txt};
\addlegendentry{MMSE, LoS (NTN-TDL-C)};
\addplot[line width=1pt, red]
table[x=snr, y=mmse_tdla, col sep=comma] {Figs/fig2c_tdl.txt};
\addlegendentry{MMSE, NLoS (NTN-TDL-A)};
\addplot[dashed, line width=1pt, blue]
table[x=snr, y=mrt_tdlc, col sep=comma] {Figs/fig2c_tdl.txt};
\addlegendentry{MRT, LoS (NTN-TDL-C)};
\addplot[dashed, line width=1pt, red]
table[x=snr, y=mrt_tdla, col sep=comma] {Figs/fig2c_tdl.txt};
\addlegendentry{MRT, NLoS (NTN-TDL-A)};
\addplot[only marks, black, mark=o]
table[x=snr, y=perf_tdlc, col sep=comma] {Figs/fig2c_tdl.txt};
\addlegendentry{MMSE, perfect CSI};
\addplot[only marks, black, mark=o, forget plot]
table[x=snr, y=perf_tdla, col sep=comma] {Figs/fig2c_tdl.txt};
\end{axis}
\node[font=\footnotesize, anchor=north] at (current axis.outer south) {(c)};
\end{tikzpicture}
\caption{Feasibility at 2.2~GHz and 600~km. (a) Spatial DoF versus aperture from~\eqref{eq:Kstar} (lines) and exact beam count (markers). (b) Rate per user in LoS from~\eqref{eq:mrt} and~\eqref{eq:mmse} (lines) and simulation (markers). (c) Rate per user under NTN-TDL-C and NTN-TDL-A with position-based (lines) and perfect-CSI (markers) precoding. 
}
\label{fig:feas}
\end{figure*}
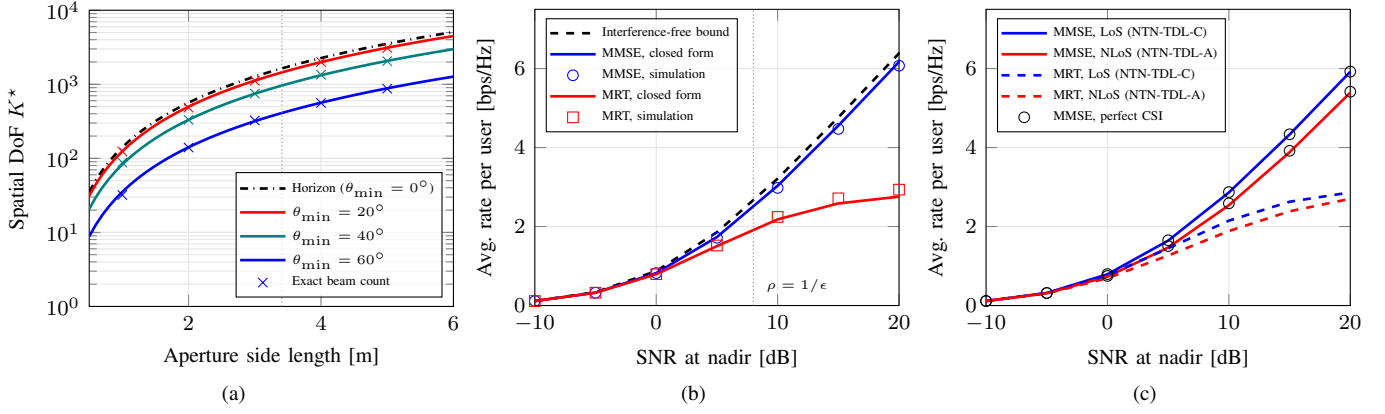

\subsection{The Spatial DoF of a Coverage Area}
In LEO satellite systems, the distance between the satellite and the user is hundreds of kilometers, several orders of magnitude larger than the extent of the local scattering around the user. Consequently, the channel of each user is well described by a single steering vector pointing in the user's direction, as reflected in the 3GPP NTN channel models~\cite{TR38811}, so that a single user sees a rank-one channel and cannot be served with spatial multiplexing, even with multiple antennas at the user. MU-MIMO, in contrast, is feasible as long as the users are far enough apart to be resolved by the satellite array, whose resolution is set by the aperture, since a larger array forms narrower beams and separates closer users. For a UPA with element spacing $d$, the steering vectors of two users are nearly orthogonal when their directions differ by one resolution beam, whose width $\lambda/(M_x d)$ in direction cosines is inversely proportional to the aperture, with $M_x$ elements per side. The coverage area seen from the satellite is bounded by the scan angle $\eta$ at its edge, which follows from the Earth radius $\RE$, the altitude $h$, and the minimum elevation angle $\thmin$. Assuming a fully digital array, so that the $M$ digital chains do not limit the result, and following the aperture--solid-angle argument of~\cite{Poon05}, the spatial DoF of the coverage equal the number of resolution beams that fit into this area, where for a planar array the area is measured in direction cosines,
\begin{equation}\label{eq:Kstar}
\begin{aligned}
\Kstar &\approx \frac{\pi\sin^2\!\eta}{\lambda^2}\,A = \pi\sin^2\!\eta\left(\frac{d}{\lambda}\right)^{2} M,
\end{aligned}
\end{equation}
where $\sin\eta = \RE\cos\thmin/(\RE+h)$,  $M=M_x^2$ is the number of antenna elements, $A=Md^2$ is the aperture area, $d$ is the element spacing, and $\lambda$ is the wavelength. Throughout this article, we use $d=\lambda/2$, for which $\Kstar\approx(\pi/4)\sin^2\!\eta\,M\le M$, approaching $(\pi/4)\big(\RE/(\RE+h)\big)^2 M$ for coverage extending to the horizon. Equivalently, $\Kstar$ is the number of eigenvalues of the channel correlation matrix over the coverage area that have similar magnitude. Fig.~\ref{fig:feas}(a) confirms that~\eqref{eq:Kstar} closely matches the exact number of orthogonal discrete Fourier transform (DFT) beams within the coverage area, and that the DoF grow with the area of the aperture. For example, a $50\times50$ array, 3.4~m across, offers about 410 DoF at $\thmin=60^\circ$, 960 at $40^\circ$, and 1450 at $20^\circ$.

\subsection{Precoding and Power: Exploiting the DoF}
In terrestrial massive MIMO with rich scattering, user channels become nearly orthogonal as the array grows, and MRT becomes close to optimal~\cite{Larsson14}. The LoS satellite channel lacks this favorable propagation, since all users lie within the coverage area, their channels span only $\Kstar$ dimensions, and with about $\Kstar$ users each beam leaks through its sidelobes into all others. We capture this by the aggregate leakage $\epsilon$, i.e., the power a user receives from all other beams relative to its own. For randomly selected users, $\epsilon$ is about 0~dB at full load, i.e., the leakage is as strong as the desired signal, whereas for spatially separated users selected by the scheduler, it drops to about $-8$~dB, dominated by the $-13$~dB first sidelobes of neighboring beams. Let $\rho$ denote the SNR of a user without interference, which includes its path loss. The leakage reaches the user through the same channel, so that $\epsilon$ depends only on the geometry. We replace the leakage of each user by its average~\cite{Ngo13} and, for MMSE, use the inverse relation between the mean-square error and the signal-to-interference-plus-noise ratio~\cite{Shi11}. The rate per user with MRT, and with MMSE for small $\epsilon$, is then approximately
\begin{subequations}\label{eq:rates}
\begin{align}
R_{\mathrm{MRT}} &\approx \log_2\Big(1+\frac{\rho}{1+\rho\,\epsilon}\Big),\label{eq:mrt}\\
R_{\mathrm{MMSE}} &\approx \log_2(1+\rho)-\log_2\big(1+b^2\epsilon\big),\label{eq:mmse}
\end{align}
\end{subequations}
where $b=\rho/(1+\rho)$. With MRT, the leakage grows with the transmit power, so that the rate saturates at $\log_2(1+1/\epsilon)$. With MMSE, the leakage costs only a small penalty, which vanishes at low SNR and tends to $\log_2(1+\epsilon)$ at high SNR, i.e., a loss of less than 1~dB. Therefore, nonlinear precoding, such as dirty-paper coding or vector perturbation, offers only a small additional gain over MMSE if the users are well separated.

Fig.~\ref{fig:feas}(b) confirms~\eqref{eq:mrt} and~\eqref{eq:mmse} for our $50\times50$ array with $\thmin=60^\circ$, i.e., $\Kstar\approx410$ users over 5~MHz. At an SNR of 20~dB at nadir, MMSE achieves 6.1~bps/Hz per user and 95\% of the interference-free bound, whereas MRT saturates at about 3~bps/Hz per user, close to the limit of 2.9~bps/Hz predicted by~\eqref{eq:mrt}. The two precoders diverge above an SNR of about 8~dB, i.e., where $\rho\epsilon\approx 1$. The total equivalent isotropic radiated power (EIRP) is fixed by the aperture and the power amplifiers, and its split over $K$ users sets the per-user SNR. Since the throughput grows linearly with the number of users but only logarithmically with the per-user SNR, the power is best spread over up to $\Kstar$ users. If the resulting per-user SNR is well above $1/\epsilon$, MMSE is needed to suppress the leakage, whereas below this SNR, MRT captures most of the MMSE throughput with a simpler precoder design.

\subsection{Is the User Position Enough?}\label{sec:pos}
Acquiring the channel state information (CSI) of hundreds of users through downlink pilots and feedback over a 600~km link is costly. Moreover, the reports are outdated by the time they arrive. Since the LoS channel is determined by the geometry, the satellite can instead compute each steering vector from the user position and its own ephemeris and attitude~\cite{Gui22}. The position is either reported by the user, who already acquires it for timing and frequency pre-compensation in NTN, or estimated from uplink reference signals such as the sounding reference signal (SRS). Such a precoder could be degraded by two effects, namely multipath and beam squint. Multipath reflections originate from scatterers within a few hundred meters of the user. Therefore, seen from the satellite, the reflections arrive from practically the same direction as the user itself, i.e., within about $0.03^\circ$, far less than the beam width of about $2^\circ$. Hence, multipath leaves the steering vector unchanged and only makes the channel frequency-selective, scaling the desired signal and the interference alike. The 3GPP NTN tapped-delay-line (TDL) models~\cite{TR38811} capture these reflections by a few delayed taps. In the LoS model NTN-TDL-C, a dominant direct path is complemented by weaker reflections. In the non-line-of-sight (NLoS) model NTN-TDL-A, in contrast, the signal arrives as a sum of reflections without a dominant path and fades according to a Rayleigh distribution.

Beam squint arises when one precoder is applied across the band, since the array response depends on frequency, so that a beam steered at the carrier frequency is slightly displaced at the band edges. The squint grows with the ratio of bandwidth to carrier frequency, with the aperture, and with the scan angle. At the coverage edge, it amounts to about 1\% of a resolution beam for 5~MHz and 10\% for 40~MHz in our case, which is negligible. Moreover, a fully digital array can remove it by computing the precoder per group of subcarriers from the geometry. Fig.~\ref{fig:feas}(c) compares the position-based MMSE precoder with MMSE based on perfect per-subcarrier CSI. At 20~dB, their rates per user differ by less than 0.5\% in both channel models, and at lower SNR by less than 2.5\%. Hence, the user position provides practically all the CSI needed for precoding. For MRT, the position-based and perfect-CSI precoders coincide, since multipath only scales the channel of each user. Relative to pure LoS, the rate per user at 20~dB decreases by 2.5\% in the LoS model and by about 11\% in the NLoS model. This loss is due to the fading of the desired signal, which no precoder can remove and which standard link adaptation handles. MRT, in contrast, saturates below 3~bps/Hz per user in both models, confirming that MMSE remains necessary at high SNR.

\section{Architecture: What Payload Does It Take?}
\label{sec:arch}

\subsection{Physical-Layer Design: How Hybrid Can the Payload Be?}
\begin{figure}[t]
\centering
\begin{tikzpicture}
\begin{axis}[
	width=8.6cm, height=6cm,
	xmin=300, xmax=2600, ymin=80, ymax=103,
	xlabel={Number of digital chains $N_{\mathrm{RF}}$},
	ylabel={Relative throughput [\%]},
	xlabel near ticks, ylabel near ticks,
	x label style={font=\footnotesize}, y label style={font=\footnotesize},
	ticklabel style={font=\footnotesize},
	scaled x ticks=false, xticklabel style={/pgf/number format/1000 sep={}},
	legend cell align=left,
	legend style={at={(0.97,0.03)}, anchor=south east},
	legend style={font=\scriptsize, inner sep=1pt, fill opacity=0.75, draw opacity=1, text opacity=1},
	grid=both, grid style={gray!20},
]
\addplot[line width=1pt, blue, mark=o, mark options={solid}]
table[x=nrf, y=pct, col sep=comma] {Figs/fig3_hybrid_th60.txt};
\addlegendentry{$\theta_{\min}=60^\circ$ ($K^\star=410$)};
\addplot[line width=1pt, teal, mark=square, mark options={solid}]
table[x=nrf, y=pct, col sep=comma] {Figs/fig3_hybrid_th40.txt};
\addlegendentry{$\theta_{\min}=40^\circ$ ($K^\star=962$)};
\addplot[line width=1pt, red, mark=triangle, mark options={solid}]
table[x=nrf, y=pct, col sep=comma] {Figs/fig3_hybrid_th20.txt};
\addlegendentry{$\theta_{\min}=20^\circ$ ($K^\star=1448$)};
\draw[densely dotted, blue] (axis cs:410,80)  -- (axis cs:410,103);
\draw[densely dotted, teal] (axis cs:962,80)  -- (axis cs:962,103);
\draw[densely dotted, red]  (axis cs:1448,80) -- (axis cs:1448,103);
\draw[dashed, black] (axis cs:300,100) -- (axis cs:2600,100);
\node[font=\tiny, anchor=south east] at (axis cs:2580,100.2) {fully digital ($M=2500$)};
\end{axis}
\end{tikzpicture}
\caption{Throughput of the beamspace hybrid payload relative to fully digital beamforming, both with MMSE, versus $N_{\mathrm{RF}}$ for the $50\times50$ array with $K=\Kstar$ users at an SNR of 20~dB. Dotted lines mark $N_{\mathrm{RF}}=\Kstar$.}
\label{fig:hyb}
\end{figure}
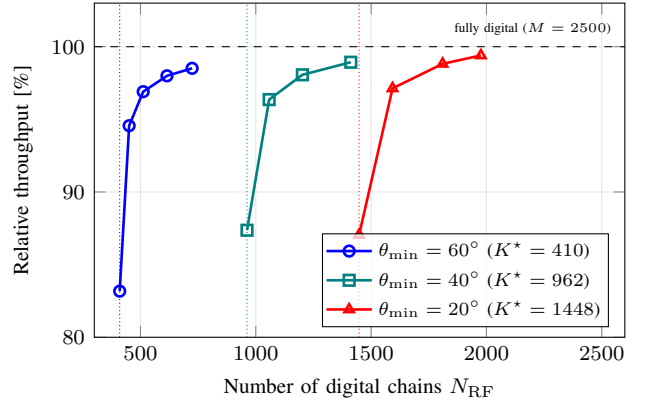

The cost and power of the digital payload are dominated by the number of digital chains, each consisting of a digital-to-analog converter and a radio-frequency (RF) chain. A fully digital array with one chain per element offers maximum flexibility, but its cost grows with the number of elements. The precoder does not need this flexibility, since the co-scheduled users occupy at most about $\Kstar$ resolvable directions. Hence, the transmit signal lies in a subspace much smaller than the array. We exploit this with a beamspace hybrid architecture, in which a fixed Fourier beamforming network forms a grid of orthogonal beams. The $N_{\mathrm{RF}}$ beams closest to the scheduled users are connected to the digital chains, and precoding is applied only across these beams. Fig.~\ref{fig:hyb} shows that, for $K=\Kstar$ users at 20~dB, the hybrid payload achieves 83--87\% of the fully digital throughput with $N_{\mathrm{RF}}=\Kstar$, and 97--99\% with $N_{\mathrm{RF}}=1.25\,\Kstar$. The additional chains are needed since users rarely lie at beam centers, so that the channel of each user spreads over several neighboring beams, in particular at the coverage edge. If the analog beams can instead be steered at the scheduled users, $N_{\mathrm{RF}}=K$ suffices, since the MMSE precoder lies in the span of their steering vectors. However, this requires adaptive phase shifters, and since an analog beam is common to the whole band, users on different resource blocks (RBs) cannot have different beams. For the fixed beam grid, the saving in chains depends on the coverage. For $\thmin=60^\circ$, 512 chains suffice instead of 2500, a fivefold reduction, whereas for $\thmin=20^\circ$, the DoF approach the horizon limit and 1810 chains are required, a reduction by only 1.4. Therefore, the digital chains should be dimensioned by the DoF of the intended coverage area rather than by the number of elements. Narrow coverage thus favors a strongly hybrid payload, whereas wide coverage pushes the design toward fully digital.

The second design aspect is where the precoder is applied relative to the orthogonal frequency-division multiplexing (OFDM) modulator of NR, which maps the data of each user onto subcarriers and converts them into time-domain samples. In frequency-domain beamforming, the precoder is applied per subcarrier, and each digital chain requires its own OFDM modulator. This allows a different user group on every RB, as needed when users are multiplexed in frequency, but the number of modulators grows with the number of chains. In time-domain beamforming, each user stream is first OFDM-modulated, and the precoder is then applied to the time-domain samples. Hence, the number of modulators grows only with the number of users. For a user group spanning the full band, both yield the same signal, except that time-domain beamforming applies one precoder across the band and thus incurs beam squint. This squint is negligible for 5~MHz (Section~\ref{sec:pos}), whereas for wider bands, frequency-domain beamforming can compensate it per subcarrier. Since the OFDM modulators and data converters dominate the digital payload~\cite{Pal22}, time-domain beamforming is preferable when the users are scheduled across the full band, whereas frequency-domain beamforming offers the flexibility needed when users are scheduled on sub-bands.

\subsection{Functional Split and Feeder Link}
The FL between the satellite and the gateway has a limited capacity, and its load depends on how the radio access network (RAN) functions are split between the satellite and the ground. For the functional splits defined by 3GPP and the O-RAN Alliance, the FL carries the user data in the form in which it crosses the split, together with control signaling. The control signaling includes the scheduling information, such as the resource allocation and the modulation and coding scheme of each user, i.e., at most about 25~Mbps for 60~bits per user in every 1-ms slot. Since it is needed for every split below the medium access control (MAC) layer, it is not compared. We consider only the precoder-related control, i.e., the precoding weights or the user positions from which the satellite computes them, since it depends on the split. Table~\ref{tab:split} compares the FL load of the main functional splits for $K=410$ users over 5~MHz with 16-bit in-phase and quadrature (I/Q) samples, ordered by increasing on-board processing.

In split~8, only the RF chains are on board. Hence, the FL must carry the time-domain samples of all 2500 elements, about 600~Gbps, which is impractical. In split~7.2, the OFDM modulation and the precoding move on board. The data plane then reduces to the frequency-domain QAM symbols of the scheduled users. With 16-bit I/Q samples, this requires about 55~Gbps, but only about 14~Gbps if each QAM symbol is sent by its bits, as in the modulation compression of O-RAN. The control signaling, in contrast, depends on where the MMSE precoder is \emph{computed}. If it is computed on the ground, the $2500\times410$ weight matrix must be forwarded at every update. With 16~bits per real and imaginary part and wideband weights updated every 10~ms, i.e., about 33~Mbit per update, this requires an FL rate of about 3~Gbps, and 25 times more if each RB carries a different user group. Moreover, the weights arrive one FL round-trip delay late. If the precoder is computed on board, it can be updated every 10~ms from the satellite ephemeris and attitude, and the control signaling reduces to the user positions. Since a user moves only a small fraction of the beam footprint within a second, its position needs to be updated only about every second, i.e., 128~bits per user or only about 50~kbps in total. In split~6, the entire physical layer (PHY) is on board, and the FL carries only the transport blocks, about 12.5~Gbps at 20~dB, whereas the MAC layer and the scheduler remain on the ground. Moving more functions on board, from split~8 toward split~6, thus reduces the FL load by more than an order of magnitude. However, it shifts processing and power consumption to the satellite, where the functions are harder to upgrade. The choice of the split therefore depends on the on-board processing and power budget and on the FL capacity. Finally, if no gateway is in view, the same traffic can be relayed over ISLs, which then carry the load of the chosen split.

\begin{table}[t]
\centering
\caption{FL load per functional split.}
\label{tab:split}
\footnotesize
\renewcommand{\arraystretch}{1.1}
\setlength{\tabcolsep}{3pt}
\begin{tabular}{@{}>{\raggedright}p{0.8cm}>{\raggedright}p{3cm}>{\raggedright}p{1.5cm}>{\raggedright\arraybackslash}p{3cm}@{}}
\noalign{\hrule height 1pt}
\textbf{Split} & \textbf{On board} & \textbf{Data} & \textbf{Control} \\
\noalign{\hrule height 0.8pt}
8 & RF & 600~Gbps & -- \\
\hline
7.2 & RF, OFDM, precoding & 14--55~Gbps & Weights, 3~Gbps \\
\hline
7.2 & RF, OFDM, precoding, precoder design & 14--55~Gbps & Positions, 50~kbps \\
\hline
6 & Full PHY & 12.5~Gbps & Positions, 50~kbps \\
\noalign{\hrule height 1pt}
\end{tabular}
\end{table}

\section{Implementation: Scheduling and Precoding}
\label{sec:impl}

\begin{table*}[!t]
\centering
\caption{Design guidelines and lessons learned for massive MU-MIMO LEO D2D payloads.}
\label{tab:rules}
\small
\renewcommand{\arraystretch}{1.15}
\begin{tabular}{@{}>{\raggedright}p{2.6cm}>{\raggedright}p{7.3cm}>{\raggedright\arraybackslash}p{7.3cm}@{}}
\noalign{\hrule height 1pt}
\textbf{Aspect} & \textbf{Guideline} & \textbf{Case study ($50\times50$)} \\
\noalign{\hrule height 0.8pt}
Spatial DoF & $\Kstar$ in~\eqref{eq:Kstar} grows with aperture and coverage & 410, 960, 1450 at $\thmin=60^\circ$, $40^\circ$, $20^\circ$ \\
\hline
Precoder & MRT at low SNR, MMSE otherwise & At 20~dB, MMSE achieves 6.1~bps/Hz per user, MRT saturates at 3~bps/Hz per user \\
\hline
CSI & User position, ephemeris, and attitude & Within 0.5\% of perfect CSI at 20~dB in NTN-TDL-C and NTN-TDL-A \\
\hline
Digital chains & $N_{\mathrm{RF}}\approx1.25\,\Kstar$ with fixed beams, $K$ with steerable beams, not $M$ & 97--99\% of fully digital, $5\times$ fewer chains at $60^\circ$ \\
\hline
Beamforming domain & Time domain for full-band users, frequency domain for sub-band users or wide bands & OFDM modulators scale with $K$ or $N_{\mathrm{RF}}$ \\
\hline
Functional split & More functions on board lower the FL load but raise payload power, same load on ISLs without a gateway in view & 600~Gbps (split 8) to 12.5~Gbps (split 6), 50~kbps control with on-board precoder \\
\hline
Timescales & Precoder every 10~ms on board, pairing every 100~ms to a few seconds & ${<}0.1\%$ loss at 10~ms, about 99\% after 5~s \\
\hline
Cells & Cell-specific beams for SSB/SIB, user data precoded across cells & $2.3\times$ mean, $5\times$ cell-edge rate \\
\noalign{\hrule height 1pt}
\end{tabular}
\end{table*}

\subsection{Precoder Update versus User Pairing}
A LEO satellite moves at about 7.6~km/s, so that every steering vector changes continuously. Two decisions must therefore follow this motion, namely \emph{how} the users are served, i.e., the precoder, and \emph{which} users are served together, i.e., the user pairing. Both depend on the user directions, but on different timescales. The precoder requires the exact directions, which change with the satellite motion, whereas the pairing depends only on the angular separation between users, which changes slowly. Fig.~\ref{fig:op}(a) shows this for static users with $K=\Kstar=410$, as a function of the user-pairing interval and for different precoder update intervals. Updating the precoder every 10~ms costs less than 0.1\% of the throughput, every 50~ms less than 0.5\%, and every 200~ms about 5\%, since the satellite then moves by more than 1~km and shifts each user direction by a noticeable fraction of a beam. Hence, updating the precoder every radio frame, i.e., 10~ms, loses practically no throughput. It is also practical, since the satellite computes the precoder from the user positions and its own ephemeris and attitude without over-the-air signaling. With such a precoder, keeping the same user pairing retains about 99\% of the throughput after 5~s and 97\% after 10~s, whereas the loss grows to 30\% after 30~s, since the selected users leave the coverage area. The user pairing, however, is more complex than the precoder design, since it requires a search over all candidate users. Therefore, it can be updated much less often, i.e., every 100~ms to a few seconds, depending on whether it is computed on board or on the ground, on the available processing, and on the FL delay.

\subsection{Cellular Control, Cell-Free Data}
\begin{figure}[t]
\centering
\pgfplotsset{fig4style/.style={
	width=8.6cm, height=5.4cm,
	xlabel near ticks, ylabel near ticks,
	x label style={font=\footnotesize}, y label style={font=\footnotesize},
	ticklabel style={font=\footnotesize},
	legend cell align=left,
	legend style={font=\scriptsize, inner sep=1pt, fill opacity=0.75, draw opacity=1, text opacity=1},
	grid=both, grid style={gray!20},
}}
\begin{tikzpicture}
\begin{axis}[fig4style,
	xmode=log, xmin=0.08, xmax=50, ymin=50, ymax=105,
	xtick={0.1,1,10,50}, xticklabels={0.1,1,10,100},
	xlabel={User-pairing interval [s]},
	ylabel={Relative throughput [\%]},
	legend style={at={(0.03,0.03)}, anchor=south west},
]
\addplot[line width=1pt, blue, mark=o, mark size=1.5pt, mark options={solid}]
table[x=T, y=p10, col sep=comma] {Figs/fig4a_pairing.txt};
\addlegendentry{Precoder updated every 10\,ms};
\addplot[line width=1pt, teal, dashed, mark=square, mark size=1.5pt, mark options={solid}]
table[x=T, y=p50, col sep=comma] {Figs/fig4a_pairing.txt};
\addlegendentry{Precoder updated every 50\,ms};
\addplot[line width=1pt, red, dotted, mark=triangle, mark size=1.8pt, mark options={solid}]
table[x=T, y=p200, col sep=comma] {Figs/fig4a_pairing.txt};
\addlegendentry{Precoder updated every 200\,ms};
\end{axis}
\node[font=\footnotesize, anchor=north] at (current axis.outer south) {(a)};
\end{tikzpicture}

\vspace{2mm}
\begin{tikzpicture}
\begin{axis}[fig4style,
	xmin=0, xmax=7.5, ymin=0, ymax=1,
	xlabel={Rate per user [bps/Hz]},
	ylabel={CDF},
	legend style={at={(0.03,0.97)}, anchor=north west},
]
\addplot[line width=1pt, gray]
table[x=fix4, y=cdf, col sep=comma] {Figs/fig4b_cdf.txt};
\addlegendentry{Fixed beams (reuse 4)};
\addplot[line width=1pt, red]
table[x=percell, y=cdf, col sep=comma] {Figs/fig4b_cdf.txt};
\addlegendentry{Per-cell pairing \& MMSE};
\addplot[line width=1pt, teal]
table[x=jointprec, y=cdf, col sep=comma] {Figs/fig4b_cdf.txt};
\addlegendentry{Per-cell pairing, cross-cell MMSE};
\addplot[line width=1pt, blue]
table[x=joint, y=cdf, col sep=comma] {Figs/fig4b_cdf.txt};
\addlegendentry{Cross-cell pairing \& MMSE};
\end{axis}
\node[font=\footnotesize, anchor=north] at (current axis.outer south) {(b)};
\end{tikzpicture}
\caption{Implementation at 2.2~GHz and 600~km. (a) Relative throughput versus user-pairing interval for static users with $K=\Kstar=410$ and different precoder update intervals. (b) Cumulative distribution function (CDF) of the rate per user for 19 cells of 64~km radius with 15 users each.}
\label{fig:op}
\end{figure}
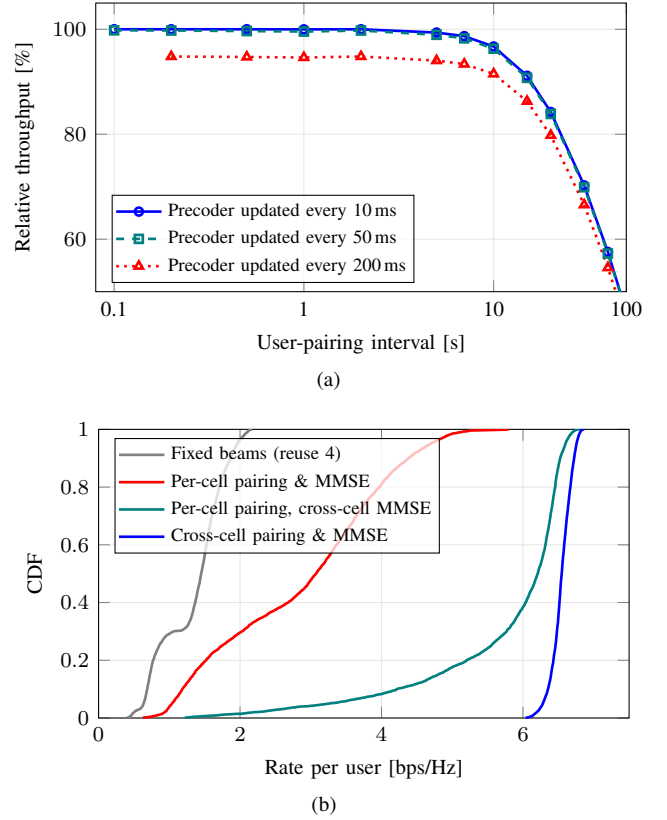

The coverage area of one satellite spans many earth-fixed NR cells, each with its own physical cell identity, synchronization signal block (SSB), and system information block (SIB). If each cell is scheduled and precoded independently on the same frequency, users near the cell borders suffer from the beams of neighboring cells, although all these cells are served by the same array and the same scheduler. We therefore propose to separate the control and data planes for the cells of one satellite, as in Fig.~\ref{fig:system}. The cell-specific signals, i.e., SSB and SIB, are transmitted on fixed cell-specific beams, which preserves legacy cell search, initial access, and mobility. These beams need not be wider than the data beams, since a wider beam requires a smaller aperture and thus loses EIRP. Instead, the narrow beams covering a cell can transmit its SSB with the same physical cell identity, e.g., by beam sweeping. The user data, in contrast, is precoded jointly across all cells of the coverage area on user-specific beams. Since all these cells are served by the same satellite, this joint precoding requires no coordination between cells. Moreover, it is transparent to the user, who only sees its user-specific demodulation reference signal (DMRS). Fig.~\ref{fig:op}(b) quantifies the benefit for 19 hexagonal cells with a radius of 64~km around nadir and 15 users per cell. Per-cell pairing and per-cell MMSE achieve an average of 2.9~bps/Hz per user, limited by inter-cell interference. Precoding jointly across cells, while keeping the per-cell pairing, raises the average to 5.8~bps/Hz and the 10th-percentile rate from 1.2 to 4.2~bps/Hz. Pairing users across cells as well yields 6.5~bps/Hz with an almost flat distribution, i.e., 2.3 times the average and five times the cell-edge rate of per-cell operation. As a baseline, tapered fixed beams with four-color frequency reuse, which serve one user per beam at a time with the same total power, achieve only 1.3~bps/Hz. Over the full coverage area, MU-MIMO with 410 users therefore achieves more than four times the throughput of such fixed beams.

\subsection{Lessons Learned and Standardization}
Table~\ref{tab:rules} summarizes the design guidelines and lessons learned in this article. Most of them are compatible with NR NTN today. Position-based precoding reuses the positions that NTN users already acquire for pre-compensation. The precoder, the hybrid architecture, and the functional split are network implementation choices, and cross-cell precoding is transparent to the user. Standard support would help to acquire the user positions efficiently and with an accuracy sufficient for precoding.

\section{Open Challenges and Future Directions}
\label{sec:future}
\textit{Uplink positioning for precoding.} Section~\ref{sec:pos} shows that the user position provides practically all the CSI needed for precoding. This position is typically obtained from the global navigation satellite system (GNSS), which may be unavailable, degraded, or spoofed. The same large array can instead estimate the direction of each user from uplink signals, such as the random-access preamble or the SRS, with a precision well below the beam width, and combine it with timing-advance and Doppler measurements into a position estimate. Since the direction of a user does not depend on frequency, this estimate directly yields the downlink steering vector, even in frequency-division duplexing without channel reciprocity. Hence, joint uplink positioning and downlink precoding is a natural step toward integrated positioning and communication.

\textit{Hotspots and multi-satellite cooperation.} The DoF in~\eqref{eq:Kstar} assume users spread over the coverage area. Traffic, however, concentrates in cities that span only a few beam footprints, so that a hotspot is limited by its local rather than the total DoF. Since several satellites are visible, multi-satellite cooperation can add DoF for hotspots, but requires tight synchronization across ISLs~\cite{Multisat24}. Whether a larger single aperture, frequency and time multiplexing within the hotspot, or cooperation among satellites is preferable remains open.

\textit{AI-native scheduling.} The user pairing in Section~\ref{sec:impl} requires a search over thousands of candidate users across dozens of cells every 100~ms to a few seconds. Learning-based schedulers, trained in a digital twin of the constellation on predictable satellite trajectories and traffic patterns, could approach the DoF at a fraction of this complexity, in line with the 3GPP framework for artificial intelligence (AI) and machine learning in the NR air interface~\cite{TR38843}.

\textit{Coexistence with terrestrial networks.} D2D services often reuse the spectrum of terrestrial mobile operators, so that the satellite must not interfere with terrestrial cells in the same band. The precoder can place nulls toward terrestrial coverage areas known from network planning, at the cost of some DoF. Quantifying this trade-off between the protected terrestrial area and the satellite throughput is an open problem for spectrum sharing.

\section{Conclusion}
\label{sec:conc}
We have shown that a single LEO satellite with a meter-scale array can serve hundreds of users on the same time-frequency resource. The aperture, the orbit altitude, and the minimum elevation angle set the spatial DoF of the coverage area, e.g., about 410 for a $50\times50$ array at $\thmin=60^\circ$. With MMSE precoding, these users achieve 6.1~bps/Hz on average at 20~dB, whereas MRT becomes interference-limited at about 3~bps/Hz. The user position and the satellite ephemeris and attitude provide channel knowledge within 0.5\% of perfect CSI at 20~dB under the 3GPP NTN channel models. Moreover, the DoF of the intended coverage area, rather than the number of elements, determine the number of digital chains of a hybrid payload. The functional split trades FL and ISL load against on-board processing. The precoder can be updated from the geometry every radio frame at a negligible loss, whereas the more complex user pairing can be kept for 100~ms to a few seconds. Finally, for the multiple NR cells under one satellite, user data can be precoded jointly across cells, while cell-specific signals remain on fixed beams. Together, these guidelines, summarized in Table~\ref{tab:rules}, increase the throughput of a single satellite more than four times compared with fixed beams and four-color frequency reuse, without new procedures at the user.

\bibliographystyle{IEEEtran}
\bibliography{refs}

\section*{Biographies}
\noindent\textsc{Bikshapathi Gouda} received the M.Tech.\ degree in communication systems from the Indian Institute of Technology Madras, India, in 2013, and the D.Sc.\ (Tech.)\ degree in wireless communications from the University of Oulu, Finland, in 2024. He was with Asarva Chips and Technologies and Cypress Semiconductors, India, working on IEEE 802.11 physical-layer algorithms, and with the University of Oulu as a Postdoctoral Researcher. Since 2026, he has been with Skylo Technologies, Espoo, Finland, where he works on physical-layer design and satellite system architecture for NTN. His research interests include wireless communications, signal processing, and optimization, with a focus on massive MIMO and NTN.

\medskip

\noindent\textsc{Neng Wang} received the Ph.D.\ degree in electrical engineering from Queen's University, Kingston, ON, Canada, in 2005. From 2007 to 2025, he was with Nortel, Qualcomm, Huawei, and Mavenir, where he led system efforts in LTE and NR modem algorithms, cellular and satellite system design, and standardization. Since 2025, he has been a Principal Wireless Systems Lead with Skylo Technologies, Espoo, Finland, where he works on systems, algorithms, and architectures for 5G and 6G satellite communications. His research interests include wireless communications, signal processing, physical- and MAC-layer design, and cellular and satellite access networks.

\medskip

\noindent\textsc{Masoud Sajadieh} received the M.A.Sc.\ and Ph.D.\ degrees in electrical and computer engineering from the University of Toronto, ON, Canada. He was with Intel and with Bell Labs, Lucent Technologies, where he led innovations in OFDM synchronization, MU-MIMO, and beamforming. He is currently a Senior Wireless Engineer with Skylo Technologies, Mountain View, CA, USA, where he works on system engineering, architecture, and signal processing for NTN. His research interests include NTN system design, MU-MIMO, and beamforming.

\medskip

\noindent\textsc{Meghna Agrawal} received the M.E.\ degree in telecommunication from the Indian Institute of Science, Bangalore, India, in 2005. She is currently Director of Advanced Network Technologies with Skylo Technologies, Mountain View, CA, USA, where she leads research and standardization on NTN. Her research interests include 3GPP NTN standardization, 5G and 6G NTN system design, and IoT over satellite.

\end{document}